\documentstyle[12pt]{article}

\def\be{\begin{equation}}
\def\ee{\end{equation}}
\def\ba{\begin{array}}
\def\ea{\end{array}}
\def\d{\partial}
\def\dps{\displaystyle}
\def\da{a_{\alpha}\frac{\partial}{\partial a_{\beta}}}
\def\db{b_{\alpha}\frac{\partial}{\partial b_{\beta}}}
\def\T{{\cal T}}

\begin{document}

\begin{titlepage}

\begin{flushright}
\end{flushright}

\vspace{3cm}

\begin{center}

{\bf \Large Free fermionic higher spin fields in $AdS_5$}

\vspace{.7cm}

K.B. Alkalaev

\vspace{.7cm}

{\em Tamm Theory Division, Lebedev Physical Institute, \\
119991 Moscow, Russia }

\vspace{3cm}

Jule 2001

\vspace{1cm}

\begin{abstract}

Totally symmetric massless fermionic fields of arbitrary spins in
$AdS_5$ are described as $su(2,2)$ multispinors. The approach is
based on the well-known isomorfism $o(4,2)\sim su(2,2)$. Explicitly
gauge invariant higher spin free actions are constructed and free
field equations are analyzed.

\end{abstract}

\end{center}

\begin{quote}
\vfill
\hrule width 5.cm
\vskip 2.mm
{\small \noindent E-mail: alkalaev@lpi.ru}
\end{quote}

\end{titlepage}

\section{Introduction}

The essential progress has recently been  achieved in the
construction of consistent interactions of massless higher spin
fields in $\rm d=5$ (for review of the theory of higher spin gauge
fields in $\rm d=2,3,4$ see \cite{V_obz,V_obz2}). It was shown in
\cite{VD5} that a consistent gravitational interaction of massless
bosonic higher spin fields in $AdS_5$ does exist at least in the
fist nontrivial order.

One of the crucial points of the work \cite{VD5} is the
description of $AdS_5$ bosonic higher spin fields by making use of
$su(2,2)$ spinor language. In particular, the $\rm 5d$ higher
spin symmetry algebra underlying such theories admits a natural
realization in terms of certain star product algebra with spinor
generating elements \cite{FL,SS,VD5}.

The aim of the present work is to elaborate the same language
being applied to $AdS_5$ higher spin fermionic fields. This is
motivated by a desire to gain a uniform description of both bosons
and fermions in order to analyze $AdS_5$ supersymmetric higher
spin gauge theories \cite{AV2}. As a first step in this direction
we formulate free fermionic action in a manifest gauge invariant
manner and analyze its general structure. We conclude that up to
irrelevant topological contributions an ambiguity in coefficients
reduces to an overall factor in front of the action of a given
spin.

One should note that fermionic higher spin fields were originally
considered in the framework of Lorentz tensor-spinors and the
corresponding free action functionals were constructed \cite{vf}.
However, the language of tensor-spinors seems to be inadequate to
the interaction problem because already at the free level
consideration one runs into great technical complications
\cite{vf}. On the other hand, as we shall see, the analysis in
terms of multispinors simplifies drastically.

The paper is organized as follows. In section 2 we collect
following \cite{VD5} some relevant facts  about
$su(2,2)$ spinor language. In section 3 we describe $\rm 5d$
fermionic higher spin fields in terms of $su(2,2)$ spinors. In
sections 4 and 5 we study linearized higher spin curvatures and
construct the quadratic fermionic action possessing all necessary
higher spin symmetries. Free field equations are analyzed in
section 6. Finally, in section 7 we give a brief summary of our
results.

%----------------------------------------------------------------------------------------
\section{Preliminaries}
%----------------------------------------------------------------------------------------

In what follows we give some formulae related to $su(2,2)$ spinors
and $AdS_5$ gravity as gauge theory of $su(2,2)$. As in
\cite{VD5}, we introduce the set of oscillators

\be
\label{ab} [a_\alpha,
b^\beta]_*=\delta_\alpha{}^\beta\;,\;\;\;[a_\alpha,a_\beta]_*=[b^\alpha,b^\beta]_*=0\;,\qquad
\alpha,\beta=1\div4\;,
\ee
where subscript $*$ refers to Weyl star-product

\be
(A*B)(a,b) = A(a,b)\:\exp(\triangle)\:B(a,b)
\ee
where

\be
\triangle = \frac{1}{2}(\frac{\overleftarrow{\d}}{\d a_\alpha}\: \frac{\overrightarrow{\d}}{\d
b^\alpha}-\frac{\overleftarrow{\d}}{\d b^\alpha}\: \frac{\overrightarrow{\d}}{\d
a_\alpha})
\ee
and commutators are taken with respect to $*$-product. One may
show that the traceless bilinears $t_\alpha{}^\beta$ of $a_\alpha$ and
$b^\beta$ commuting with $a_\alpha b^\alpha$  span $sl_4({\bf
C})$ algebra \cite{VD5,FL}. The $su(2,2)$ real form of $sl_4({\bf
C})$ results from the reality condition

\be
\bar{a}_\alpha = b^\beta C_{\beta\alpha}\;,\qquad \bar{b}^\alpha =
C^{\alpha\beta}a_\beta\;,
\ee
where bar denotes the complex conjugation  while $C$ are some real
antisymmetric matrices satisfying

\be
C_{\alpha\gamma}C^{\beta\gamma}=\delta_\alpha{}^\beta\;.
\ee
The oscillators $b^\alpha$ and $a_\alpha$ form the fundamental and the
conjugated fundamental representations of $su(2,2)$ equivalent to
the two spinor representations of $o(4,2)$.

The $AdS_5$ gravitational fields are identified with the connection
1-forms taking values in the $AdS_5$ algebra $su(2,2)$:

\be
\label{gf} \Omega = \Omega^\alpha{}_\beta\:t_\alpha{}^\beta\;. \ee
The connection should decompose as the usual Lorentz connection
and the frame field. We may perform this split in a
gauge-invariant manner by introducing a compensator field
\cite{SW,VP,VD5} which is antisymmetric bispinor
$V^{\alpha\beta}=-V^{\beta\alpha}$. The compensator is normalized
so that $V_{\alpha\gamma}V^{\beta\gamma}=\delta_\alpha{}^\beta$
and
$V_{\alpha\beta}=1/2\varepsilon_{\alpha\beta\gamma\rho}V^{\gamma\rho}$.
The Lorentz subalgebra is then identified with its
stability algebra. This allows one to define the frame field
and Lorentz connection as follows \cite{VD5}

\be
\ba{l}
\dps E^{\alpha\beta} = DV^{\alpha\beta}\equiv dV^{\alpha\beta}
+\Omega^\alpha{}_\gamma V^{\gamma\beta} +\Omega^\beta{}_\gamma
V^{\alpha\gamma}\;,\;\;\;\; E^{\alpha\beta} V_{\alpha\beta}=0\;,
\\
\\
\dps W^\alpha{}_\beta = \Omega^\alpha{}_\beta
+\frac{1}{2}E^{\alpha\gamma}V_{\gamma\beta}\;.
\\

\ea
\ee
where we omitted an explicit dependence on the cosmological
parameter $\lambda$. The $AdS_5$ field strength corresponding to
gauge fields (\ref{gf}) has the form

\be
R_\alpha{}^\beta= d\:\Omega_\alpha{}^\beta
+\Omega_\alpha{}^\gamma\wedge\Omega_\gamma{}^\beta
\ee
and the background $AdS$ space is extracted by the zero-curvature
condition $R_\alpha{}^\beta =0$ \cite{V_obz,V_obz2}. The corresponding background
frame field and Lorentz connection will be denoted
$h^\alpha{}_\beta$ and $w_0{}^\alpha{}_\beta$.

We will treat $V^{\alpha\beta}$ as a symplectic form that allows
to rise and lower spinor indices in the Lorentz covariant way
$A^\alpha=V^{\alpha\beta}A_\beta\;,\;A_\alpha=A^\beta
V_{\beta\alpha}\;$.

In the subsequent sections we will make use of this spinor
language to describe $AdS_5$ higher spin fermionic fields as
$su(2,2)$ multispinors. It will be shown that fermionic massless
fields corresponding to the totally symmetric $\gamma$-transverse
tensor-spinor representations of Wigner little group $SO(3)$ are
appropriately described as $su(2,2)$ traceless multispinors
$\Omega^{\alpha(m)}_{\beta(n)}$ with $|m-n|=1$. When handling with
multispinors we adhere conventions introduced and explained in
\cite{V1}. Namely, a number of symmetric indices is indicated in
parentheses.  Lower and upper indices denoted by the same letter
are assumed to be symmetrized separately. After this
symmetrization is carried out, the maximal possible number of
lower and upper indices denoted by the same letter should be
contracted.

%----------------------------------------------------------------------------------------
\section{Fermionic fields}
%----------------------------------------------------------------------------------------

In Ref. \cite{vf} fermionic $\rm AdS$ massless higher spin fields
corresponding to the totally symmetric tensor-spinor irreducible
representations of the flat little group $SO(3)$ are described as
Lorentz tensor-spinors. In this section we present the formulation
in terms of $su(2,2)$ multispinors. We exploit the fact that all
required Lorentz higher spin tensors emerge under dimensional
reduction of the particular representation of $o(4,2)$
\cite{BMV,VD5}.

According to \cite{vf} in order to describe a $\rm 5d$ massless
field of spin $s^\prime = s+3/2$ a set of 1-forms
$dx^{\underline{n}}\omega_{\underline{n}}{}^{a(s),\:b(t),\;\alpha}$
is to be introduced with fixed $s$, all $0\leq t\leq s$, where $\alpha =
1\div 4$ is Dirac spinor index, $a,b = 0\div 4 $ are Lorentz
vector indices and underlined indices $\underline{n}=0\div 4$ are
used for differential forms and vectors in $\rm 5-$dimensional
space-time. These tensors represent traceless Young diagrams with
respect to Lorentz vector indices and satisfy the
$\gamma$-transversality condition. The field
$\omega_{\underline{n}}{}^{a(s),\:b(0),\;\alpha}$ is treated as
dynamical spin $s^\prime$ field. All other fields with $t>0$ serve
as auxiliary ones which are needed for constructing of consistent
higher spin curvatures but will be shown to fall out of the free
actions.

The collection of the higher spin 1-forms with all $0\leq t\leq s$
can be interpreted  as a result of the dimensional reduction  of
the irreducible tensor-spinor representation of $AdS_5$ algebra
$o(4,2)$ described by a  traceless two-row  Young diagram
$\Omega^{\;A(s),\:B(s),\;\hat{\alpha}}\equiv
dx^{\underline{n}}\Omega_{\underline{n}}{}^{\;A(s),\:B(s), \;
\hat{\alpha}}\;\;$, $\;A,B=0\div 5\;$,$\;\hat{\alpha}=1\div 8$,
which is subject to the $\Gamma$-transversality condition. Every
spin appears twice since under dimensional reduction $\rm 6d$
Dirac spinor splits into two $\rm 5d$ spinors.

Now let us examine identification of this
rectangular representation with its $su(2,2)$ multispinor
counterpart. This schematically looks as follows

\be
\Omega^{\alpha(s+1)}_{\beta(s)}\oplus \stackrel{-}{\Omega}{}^{\alpha(s)}_{\beta(s+1)}
 \sim \Omega^{\;A(s),\;B(s), \; \hat{\alpha}}\;,
\ee
where $\alpha,\beta = 1\div4$ and the mutually conjugated
multispinors are assumed to be traceless:

\be
\Omega^{\alpha(s)\gamma}_{\beta(s-1)\gamma} =
\;\stackrel{-}{\Omega}{}^{\alpha(s-1)\delta}_{\beta(s)\delta} =0\;.
\ee
The simplest way to establish this isomorfism is to compare the
dimensionalities. The straightforward calculation shows that the
number of independent components of $o(4,2)$ spin-tensor is

\be
n(s) = \frac{5}{3}(s+1)(s+2)^{3}(s+3)\;.
\ee
Similarly, one finds that

\be
{\rm dim}\:\Omega^{\alpha(s+1)}_{\beta(s)}={\rm dim}\:\stackrel{-}{\Omega}{}^{\alpha(s)}_{\beta(s+1)}=\frac{n(s)}{2}.
\ee
To decompose multispinor $\Omega^{\alpha(s+1)}_{\beta(s)}$ in
irreducible Lorentz tensors we make use of the compensator
$V_{\alpha\beta}$

\be
\label{L} \Omega^{\alpha(s+1)}_{\beta(s)} =
\sum_{t=0}^{s}\omega^{\alpha(s+1)\gamma(t),\gamma(s-t)}
V_{\beta(s),\gamma(s)}\;,
\ee
where $V_{\alpha(m),\:\beta(m)}$ denotes
$V_{\alpha_1\beta_1}V_{\alpha_2\beta_2}\ldots V_{\alpha_m\beta_m}$. The similar
decomposition holds for conjugated multispinor
$\stackrel{-}{\Omega}{}^{\alpha(s)}_{\beta(s+1)}$. Lorentz higher
spin fields $\omega^{\alpha(s+t+1), \beta(s-t)}$, $0\leq t\leq
s$ represent $su(2,2)$ Young diagram and all traces with
$V^{\alpha\beta}$ are zero:

\be
\label{ic}
\omega^{\alpha(s+t+1),\:\alpha \beta(s-t-1)} =0\;,\qquad
\omega^{\alpha(s+t+1),\: \beta(s-t)}V_{\alpha\beta} =0\;.
\ee
It is easy to get convinced that Lorentz higher spin tensors
$\omega(t,s)$ in vector and spinor notations have the same number
of independent components:

\be
\ba{l}
\label{dims}

{\rm dim}\:\omega^{a(s),\:b(t),\;\alpha} =
{\rm dim}\:\omega^{\alpha(s+t+1),\:\beta(s-t)}
\\
\\
=\dps\frac{10}{3}(s+2)(s+t+3)(s-t+1)(t+1)\;.
\ea
\ee
Traceless multispinors with equal number of the upper and lower
indices $\Omega^{\alpha(s)}_{\beta(s)}$ correspond to bosonic
fields with spin $s^\prime = s+1$ as was demonstrated in
\cite{SS,VD5}. In \cite{AV2} we show that fermionic higher
spin gauge fields discussed in this paper along with bosonic ones considered in
\cite{SS,VD5} originate from the HS algebra related to special
Heisenberg-Clifford superalgebra \cite{FL,Vc}.

%----------------------------------------------------------------------------------------
\section{Higher spin linearized curvatures}
%----------------------------------------------------------------------------------------

In accordance with the analysis of the previous section, $\rm 5d$
higher spin $s+3/2$ fields can be described by 1-forms
$\Omega^{\alpha(s+1)}_{\beta(s)}(x)\equiv
dx^{\underline{n}}\:\Omega_{\underline{n}\;}{}^{\alpha(s+1)}_{\beta(s)}(x)$
and their conjugates which are traceless multispinors symmetric in
the upper and lower indices. In what follows we exploit the
formalism of generating function of oscillators (\ref{ab}) which
is convenient for handling with arbitrary-rank tensors and have
often been used for description of higher spin fields. As in
\cite{VD5}, we introduce

\be
\label{gen} \Omega(a, b|x) =
\Omega^{\alpha(s+1)}_{\beta(s)}(x)\:a_{\alpha(s+1)}b^{\beta(s)}
\;.
\ee
The corresponding $\rm 5d$ linearized higher spin field strength has the form

\be
R = d\Omega+\Omega_0*\wedge\Omega +\Omega*\wedge\Omega_0
\ee
with

\be
\Omega_0(a,b|x)=\Omega_0{}^\alpha{}_\beta\; a_\alpha b^\beta,\qquad \Omega_0{}^\alpha{}_\alpha
=0\;,
\ee
which represents background gravitational fields satisfying the zero-curvature condition
$d\Omega_0+\Omega_0\wedge*\Omega_0=0$ (for more details see \cite{V_obz,V_obz2}). Equivalently, using (\ref{ab})

\be
\label{curv}
R(a,b|x) = d\:\Omega(a,b|x) +
\Omega_0{}^\alpha{}_\beta (b^\beta\frac{\d}{\d b^\alpha}-a_\alpha\frac{\d}{\d
a_\beta})\wedge \Omega(a,b|x)\;.
\ee
The component form of the curvatures (\ref{curv}) is

\be
\label{curvcomp}
R^{\;\alpha(s+1)}_{\;\beta(s)} =
d\:\Omega^{\;\alpha(s+1)}_{\;\beta(s)}-(s+1)\:\Omega_0{}^\alpha{}_\gamma\:
\wedge \Omega^{\;\gamma\:\alpha(s)}_{\;\beta(s)}-
s\:\Omega_0{}^\gamma{}_\beta\:
\wedge \Omega^{\;\alpha(s+1)}_{\;\beta(s-1)\:\gamma}\:.
\ee
The linearized (abelian) higher spin transformation are

\be
\label{hstr}
\delta\Omega(a,b|x)=D_0\epsilon(a,b|x)\;,
\ee
where the background covariant derivative has the form

\be
D_0=d +\Omega_0{}^\alpha{}_\beta (b^\beta\frac{\d}{\d
b^\alpha}-a_\alpha\frac{\d}{\d a_\beta})\;.
\ee
It is easy to see that $\delta R(a,b|x)=0$.

In what follows  we will use the two sets of the differential
operators  in the spinor variables introduced in \cite{VD5}:

\be
\label{S}
S^- = a_\alpha \frac{\d}{\d b^\beta}V^{\alpha\beta}\:,\qquad
S^+ = b^\alpha \frac{\d}{\d a_\beta} V_{\alpha\beta}\:,\qquad
\dps S^0 = N_b - N_a\:,
\ee
and
\be
\label{T}
T^- = \frac{1}{4} \frac{\partial^2 }{ \partial a_\alpha
\partial b^\alpha }\:,\qquad T^+ = a_\alpha b^\alpha \:,\qquad
T^0 = \frac{1}{4}( N_a +N_b  +4 )\;,
\ee
where

\be
\label{N} N_a=a_\alpha\frac{\d}{\d a_\alpha} \qquad {\rm and}
\qquad N_b=b^\alpha\frac{\d}{\d b^\alpha}\;.
\ee
With the aid of (\ref{S}) and (\ref{T}) one can easily reformulate
all irreducibility conditions on $\Omega(a,b)$ in the form:

\be
T^- \Omega(a,b)=0\;,\qquad (1+S^0)\Omega(a,b) =0\;.
\ee
As demonstrated in section 3, the higher spin gauge field
$\Omega$ admits the representation in terms of the Lorentz -
irreducible higher spin fields (\ref{L}). Here we take advantage
of introducing of operators (\ref{S}) and rewrite (\ref{L}) as follows:

\be
\label{expan}
\Omega(a,b|x) = \sum_{t=0}^{s} (S^+)^t \:\omega^t(a,b|x),
\ee
where

\be
\omega^t(a,b|x) = \omega^{\alpha(s+t+1),\; \beta(s-t)}\:(x) a_{\alpha(s+t+1)} b_{\beta(s-t)}.
\ee
All irreducibility conditions on the Lorentz multispinors
$\omega^t(a,b)$, namely Young diagram property and V-transversality (\ref{ic}),
can be rewritten as:

\be
\label{irr}
S^-\omega^t(a,b) = 0\;, \qquad T^-\omega^t(a,b) = 0\;.
\ee
The linearized higher spin curvatures (\ref{curv}) admit the
expansion analogous to (\ref{expan}):

\be
\label{cr2} R(a,b|x) = \sum_{t=0}^s (S^+)^t \: r^t(a,b|x)\:,
\ee
with $r^t(a,b)$ satisfying (\ref{irr}). One can make sure that the
higher spin gauge invariance of (\ref{cr2}) requires for $r^t$ the following
structure \cite{vf}

\be
\label{taus}
r^t = {\cal D} \omega^t  +{\cal T}^-\omega^{t+1} +{\cal T}^0\omega^t
+{\cal T}^+\omega^{t-1}
\ee
with the corresponding gauge transformation which leave Lorentz higher
spin curvatures (\ref{taus}) invariant

\be
\label{taus2}
\delta\omega^t = {\cal D} \epsilon^t  +{\cal T}^-\epsilon^{t+1} +{\cal T}^0\epsilon^t
+{\cal T}^+\epsilon^{t-1}
\ee
Here 0-forms $\epsilon^t$ are gauge parameters. ${\cal D}$ is the
background Lorentz covariant derivative

\be
\label{LD}
{\cal D} =d+w_0^{\alpha}{}_{\beta}(\da+\db)\;,
\ee
and operators ${\cal T}^-$, ${\cal T}^+$ and ${\cal T}^0$ have the form:

\be
\label{T+}
{\cal T}^+ =-(1+S^0)\;h^{\alpha}{}_{\beta}a_{\alpha}\frac{\partial}{\partial
b_{\beta}}
\ee

\be
\label{T0}
{\cal T}^0=-\frac{1}{S^0}\; h^{\alpha}{}_{\beta}(\db - \da
+\frac{2}{S^0-2}(b_\gamma\frac{\d}{\d a_\gamma})a_{\alpha}\frac{\partial}{\partial
b_{\beta}})
\ee

\be
\label{T-}
\ba{c}
\dps{\cal T}^- = \frac{1}{(S^0)^2}\;h^{\alpha}{}_{\beta}((2-S^0)b_{\alpha}\frac{\partial}{\partial
a_{\beta}} + b_\gamma\frac{\d}{\d a_\gamma}(\db - \da)
\\
\\
\dps+
\frac{1}{S^0-3}\;(b_\gamma\frac{\d}{\d a_\gamma})^{2}a_{\alpha}\frac{\partial}{\partial
b_{\beta}})

\ea
\ee

These operators satisfy the following important relations

\be
\ba{l}
\dps \{\T^0,\T^- \}=\{\T^0,\T^+\} = 0,
\\
\\
(\T^-)^2=0\;,\;(\T^+)^2=0\;,
\\
\\
\{\T^-,\T^+\}+(\T^0)^2 + {\cal D}^2 =0\;.
\\

\ea
\ee
Note that coefficients in (\ref{T+}-\ref{T-}) can be changed by
field redefinitions of the form
$\widetilde{\omega}^t=C(t,s)\omega^t$ with $C\neq 0$.

It is worth noting that the operator ${\cal T}^0$ in (\ref{taus2})
may be viewed as a higher spin generalization of the operator
$\gamma_\mu$ in the super-AdS transformation for gravitino
field $\delta\psi_\mu={\cal D}_\mu \epsilon +
i\gamma_\mu\epsilon$ \cite{ads}.

%----------------------------------------------------------------------------------------
\section{Higher spin fermionic action}
%----------------------------------------------------------------------------------------

The aim of this section is to formulate the action principle for
the free gauge massless fermionic fields in $AdS_5$. We exploit the
same technique developed previously for construction of $\rm 5d$
bosonic actions \cite{VD5}.

As in \cite{vf,VD5}, actions for massless
fermionic fields with spin $s^\prime = s+3/2$ are assumed to be of the form

\be
\label{act}
{\cal S}^{s+3/2}_2 =\int_{{\cal M}^5} \hat{H}\wedge  R^s(a_1,b_1)\wedge
\stackrel{-}{R}{}^s(a_2,b_2)|_{a_i=b_i=0}\;,
\ee
where $R^s$ denotes linearized
higher spin curvatures (\ref{curv}) constructed in the previous
section and 1-form $\hat{H}$ is the following differential
operator

\be
\label{H}
\ba{l}
\dps \hat{H}=\zeta(p,q) h_\alpha{}^\beta \frac{\d^2}{\d a_{2\alpha} \d b_1^\beta}\hat{c}_{12}\hat{c}_{12}
+\gamma(p,q) h_\alpha{}^\beta \frac{\d^2}{\d a_{1\alpha} \d b_2^\beta}
\\
\\
\dps+\alpha(p,q) h_{\alpha\beta} \frac{\d^2}{\d a_{1\alpha} \d
a_{2\beta}}\hat{b}_{12}\hat{c}_{12}
+\beta(p,q) h^{\alpha\beta} \frac{\d^2}{\d b_1^\alpha \d
b_2^\beta}\hat{a}_{12}\hat{c}_{12}\;.
\\

\ea
\ee
Here $h_{\alpha}{}^{\beta}$ is the background frame field and the
coefficients $\alpha,\;\beta,\;\gamma$ and $\zeta$ responsible for
various types of indices contractions between the frame field and
curvatures depend on the operators:

\be
p=\hat{a}_{12}\hat{b}_{12}\;,\qquad q=\hat{c}_{12}\hat{c}_{21}\;,
\ee
where

\be
\label{abg}
\ba{cc}
\dps\hat{a}_{12} = V_{\alpha\beta}\frac{\d^2}{\d a_{1\alpha} \d
a_{2\beta}}\;,&
\qquad\dps \hat{b}_{12} = V^{\alpha\beta}\frac{\d^2}{\d b_1^\alpha \d
b_2^\beta}\;,
\\
\\
\dps\hat{c}_{12} = \frac{\d^2}{\d a_{1\alpha} \d b_2^\alpha}\;,&
\dps\hat{c}_{21} = \frac{\d^2}{\d a_{2\alpha} \d b_1^\alpha}\;.
\\
\\
\ea
\ee
The action is invariant under complex conjugation $\bar{{\cal
S}}_2={\cal S}_2$ regardless of the particular form of the real coefficients
$\alpha,\;\beta,\;\gamma$ and $\zeta$.

It is clear that generally the action (\ref{act}) does
not describe massless higher spin fields because of too many
dynamical variables associated with extra fields $\omega^{s,\:t}$
with $t>0$. In order to eliminate the corresponding degrees of
freedom one should fix the operator $\hat{H}$ in an appropriate
way. The leading principle for the determination of the form of
$\hat{H}$ is the {\it extra field decoupling condition} \cite{vf}
which requires the variation of the quadratic action with respect
to the extra fields to vanish:

\be
\label{exdc} \frac{\delta {\cal S}_2^{s+3/2}}{\delta\omega^{s+3/2,\:t}} \equiv 0\;, \qquad {\rm
for}\;\; t>0\;.
\ee
The counting of physical degrees of freedom after complete gauge
fixing performed in section 6 demonstrates that such an action
does describe a massless spin $s+3/2$ field.

Note that there is an ambiguity in coefficients $\alpha(p,q)$ and
$\beta(p,q)$ due to the freedom in adding the surface terms

\be
\ba{c}
\delta{\cal S}_2 =
\dps \int_{{\cal M}^5} d(\Phi (p,q)\hat{c}_{12} R(a_1,b_1)\wedge
\stackrel{-}{R}(a_2,b_2)|_{a_i=b_i=0})
\\
\\
\dps \int_{{\cal M}^5} \frac{\d \Phi (p,q)}{\d p}(h_{\alpha\beta} \frac{\d^2}{\d a_{1\alpha} \d
a_{2\beta}}\hat{b}_{12}\hat{c}_{12}
-h^{\alpha\beta} \frac{\d^2}{\d b_1^\alpha \d b_2^\beta}\hat{a}_{12}\hat{c}_{12})
\\
\\\dps\wedge R(a_1,b_1)\wedge\stackrel{-}{R}(a_2,b_2)|_{a_i=b_i=0}
\\

\ea
\ee
where, using the manifest $su(2,2)$ invariance of our formalism, we
replace the differential $d$ with the background covariant
derivative and make use of the Bianchi identities $D_0(R)=0$. As a
result, the variation of the coefficients of the form

\be
\delta\alpha(p,q) = \epsilon(p,q)\;,\qquad  \delta\beta(p,q) = - \epsilon(p,q)
\ee
does not affect the physical content of the action, {\it i.e.} the
combination $\alpha(p,q) +\beta(p,q)$ has invariant meaning.

Let us now consider the extra field decoupling condition.
Since the generic variation of the  linearized fermionic
curvatures is $\delta R= D_0\delta\Omega$ and because the action is formulated in the
$AdS_5$ covariant way, integrating by parts one obtains the
generic variation of ${\cal S}_2$:

\be
\label{var} \delta{\cal S}_2 = \int_{{\cal M}^5} D_0\hat{H}\wedge
\delta\Omega(a_1,b_1)\wedge \stackrel{-}{R}(a_2,b_2))|_{a_i=b_i=0}\;\;\;
+\;\;\;{\rm c.c.\;\; part}\;.
\ee
The derivative $D_0$ produces the frame field every time it meets
the compensator. Taking into account $D_0 h^{\alpha\beta}
=0,\;h^\alpha{}_{\beta} =h^{\alpha\gamma}V_{\gamma\beta}$ and
denoting $H^{\alpha\beta}=H^{\beta\alpha} =
h^\alpha{}_\gamma\wedge h^{\beta\gamma}$ one finds

\be
\label{varH}
\ba{l}
D_0\hat{H}=\dps \rho_1 H^\beta{}_\alpha \frac{\d^2}{\d a_{2\alpha} \d b_1{}^\beta }
\hat{c}_{12}\hat{c}_{12}+\rho_2 H^\beta{}_\alpha \frac{\d^2}{\d a_{1\alpha} \d b_2{}^\beta }
\\
\\
\dps +\rho_3 H_{\alpha\beta} \frac{\d^2}{\d a_{1\alpha} \d a_{2\beta}}
\hat{b}_{12}\hat{c}_{12}
 +\rho_3 H^{\alpha\beta} \frac{\d^2}{\d b_1^\alpha \d b_2^\beta}
\hat{a}_{12}\hat{c}_{12}\;,
\\

\ea
\ee
where

\be
\label{rhos}
\ba{l}
\dps\rho_1 =(1+p\frac{\d}{\d p})(-2\zeta + (\alpha+\beta)) ,
\\
\\
\dps\rho_2 = (1+p\frac{\d}{\d p})(-2\gamma + q(\alpha+\beta)),
\\
\\
\dps\rho_3= \frac{\d }{\d p}(\gamma - q \zeta).

\ea
\ee
According to (\ref{expan}) the extra fields variation has the form

\be
\label{ex}
\delta\Omega^{ex}(a,b)=S^+\xi(a,b)
\ee
with an arbitrary $\xi(a,b)$ satisfying $(N_a-N_b-3)\xi(a,b)=0$.
Thereby, the extra field decoupling condition (\ref{exdc}) amounts
to

\be
\label{eq2}
\ba{l}
\dps\frac{\d \rho_2}{\d p} - \frac{\d \rho_2}{\d q} +\rho_3 =0 ,
\\
\\
\dps\frac{\d \rho_3}{\d p} - \frac{\d \rho_3}{\d q} =0 ,
\\
\\
\dps\rho_1 +\rho_3 =0.

\ea \ee
For the trivial solution $\rho_i=0$ the covariant derivative of
$\hat{H}$ vanishes, $D_0 \hat{H}=0$, and the corresponding action
functional is a topological invariant. From (\ref{rhos}) it
follows that $\rho_i=0$ provided that

\be
\label{top}
\ba{l}
(\alpha+\beta)^{\rm top} = 2\zeta^{\rm top}\;,
\\
\gamma^{\rm top} = q\zeta^{\rm top}\;,

\ea
\ee
where the function $\zeta^{\rm top}$ is an arbitrary polynomial of
degree $(s-1)$ with respect to $q$ and $p$. Clearly, by adding of
topological invariants with the coefficients satisfying
(\ref{top}) we can always set $\zeta = 0$ in (\ref{act}-\ref{H}).
Since the coefficients $\alpha,\;\beta$ and $\gamma$ are
assumed to be of the form $\sum_{k=0}^{m}\upsilon^{(k,\:m)}p^k
q^{m-k}$ where $\upsilon^{(k,\:m)}$ are real numbers ($m=s-1$ for
$\alpha,\beta$ and $m=s$ for $\gamma$ ), we readily write
down the general solution  to (\ref{eq2}) (modulo topological
invariants) as

\be
\zeta=0\;,
\ee
\be
\label{gamma}
\gamma(p+q)= \gamma^{(s)}\:(p+q)^s\;,
\ee
where the real coefficient $\gamma^{(s)}$ is arbitrary, and

\be
\label{alpha}
(\alpha+\beta)(p,q) =-s\gamma^{(s)}\int_0^1
d\tau(p\tau+q)^{s-1}\;.
\ee
An overall factor $\gamma^{(s)}$ in front of the action of a given spin
cannot be fixed from the analysis of the free action and represents the
leftover ambiguity in the coefficients (besides the trivial ambiguity (42)).

In conclusion of this section let us write down the component form
of the action

\be
\label{actcomp}
\ba{l}
\dps {\cal S}_2^{s+3/2}=\dps\gamma^{(s)}\sum_{k=0}^{s}\int_{{\cal M}^5}A^k_s\;
h^\gamma{}_\rho\:R^{\rho\:\alpha(s-k)\:\sigma(k)}_{\beta(s-k)\:\delta(k)}\:
\stackrel{-}{R}{}_{\gamma\:\alpha(s-k)\:\mu(k)}^{\beta(s-k)\:\nu(k)}\:
V_{\sigma(k),\:\nu(k)}V^{\delta(k),\:\mu(k)}
\\
\\
\dps-\gamma^{(s)}\sum_{k=0}^{s-1}\int_{{\cal M}^5}B^k_{s-1}
h_{\gamma\rho} R^{\gamma\:\alpha(s-k)\:\sigma(k)}_{\beta(s-k-1)\:\delta(k+1)}

\stackrel{-}{R}{}_{\alpha(s-k)\:\mu(k+1)}^{\rho\:\beta(s-k-1)\:\nu(k)}\:
V_{\sigma(k),\:\nu(k)}V^{\delta(k+1),\:\mu(k+1)}\;\;,
\\
\\
\ea
\ee
where

\be
A^k_s =((s+1)!)^2\frac{s!}{k!(s-k)!}\;, \quad
B^k_{s-1} =s^2((s+1)!)^2\frac{s!}{(k+1)!(s-k-1)!}\;.
\ee
The action (\ref{actcomp}) is by construction invariant under the
linearized gauge transformations (\ref{hstr}) as follows from
the invariance of the linearized curvatures (\ref{curvcomp}).

%----------------------------------------------------------------------------------------
\section{Free field equations}
%----------------------------------------------------------------------------------------

In this section we derive free field equations. Taking account of
(\ref{varH}-\ref{rhos}) and (\ref{gamma}-\ref{alpha}) we rewrite
the nontrivial part of the variation (\ref{var}) as follows

\be
\ba{l}
\label{var2} \delta{\cal S}_2^{s+3/2} =
\\
\\
\dps-s \gamma^{(s)}\int_{{\cal M}^5}(p+q)^{s-1}
((2\frac{s+1}{s}p+\frac{s+2}{s}q) H^\beta{}_\alpha \frac{\d^2}{\d
a_{1\alpha} \d b_2{}^\beta }+ H^\beta{}_\alpha \frac{\d^2}{\d
a_{2\alpha} \d b_1{}^\beta } \hat{c}_{12}\hat{c}_{12}
\\
\\
\dps -H_{\alpha\beta} \frac{\d^2}{\d a_{1\alpha} \d a_{2\beta}}
\hat{b}_{12}\hat{c}_{12}
 - H^{\alpha\beta} \frac{\d^2}{\d b_1^\alpha \d b_2^\beta}
\hat{a}_{12}\hat{c}_{12})

\wedge\delta\omega^0(a_1,b_1)\wedge \stackrel{-}{r}{}^0(a_2,b_2)\;+\;{\rm c.c.}\;.
\ea
\ee
By virtue of Young symmetry properties of the dynamical field and its
curvature the variation (\ref{var2}) takes the form:

\be
\ba{l}
\label{var3} \delta{\cal S}_2^{s+3/2} =
\dps -\gamma^{(s)}\int_{{\cal M}^5}((p+q)^{s-1}
((2s+2)p+(s+2)q)H^\beta{}_\alpha \frac{\d^2}{\d
a_{1\alpha} \d b_2{}^\beta}
\\
\\
\dps+2s H^\beta{}_\alpha \frac{\d^2}{\d
a_{2\alpha} \d b_1{}^\beta } \hat{c}_{12}\hat{c}_{12})

\wedge
\delta\omega^0(a_1,b_1)\wedge \stackrel{-}{r}{}^0(a_2,b_2))\;
+\;{\rm c.c.}\;.
\\
\\

\ea
\ee
Substituting

\be
\ba{l}
\omega^0(a,b) = \omega^{0\;\alpha(s+1),\;\beta(s)}a_{\alpha(s+1)}b_{\beta(s)}
\\
\\
\stackrel{-}{r}{}^0(a,b) = \stackrel{-}{r}{}^0_{\;\beta(s+1),\;\alpha(s)}b^{\beta(s+1)}a^{\alpha(s)}
\ea
\ee
one finds

\be
\ba{c}
\label{var4} \delta{\cal S}_2^{s+3/2} \sim
\dps \int_{{\cal M}^5}((3s+2)(s+1)H^\gamma{}_\alpha\:\wedge\:\stackrel{-}{r}{}^0_{\gamma\:\alpha(s),\;\beta(s)}
-s(3s+4)H^\gamma{}_\beta\:\wedge\:\stackrel{-}{r}{}^0_{\alpha(s+1),\;\beta(s-1)\:\gamma})
\\
\\
\wedge \delta\omega^{0\:\alpha(s+1),\:\beta(s)}\;\;+\;\;{\rm c.c.}\;=0.
\\
\\

\ea
\ee
The equations of motion resulting from the variation
(\ref{var4}) have the form

\be
\label{EM}
\ba{l}
\dps(s+1)H^\gamma{}_\alpha\:\wedge\:\stackrel{-}{r}{}^0_{\gamma\:\alpha(s),\;\beta(s)}-
s\frac{s+2}{3s+2}H^\gamma{}_\beta\:\wedge\:\stackrel{-}{r}{}^0_{\alpha(s+1),\;\beta(s-1)\:\gamma}
\\
\\
\dps +2s(s+1)\frac{s+1}{3s+2}H^\gamma{}_\alpha\:\wedge\:\stackrel{-}{r}{}^0_{\alpha(s)\:\beta,\:\beta(s-1)\:\gamma}=0.
\\
\\
\ea
\ee
and the analogous equations for the conjugated field.
These are invariant under the gauge transformation of the form (\ref{taus2})

\be
\label{gtr}
\ba{l} \dps \delta\stackrel{-}{\omega}{}^0_{\alpha(s+1),\:\beta(s)} =
{\cal D}\stackrel{-}{\epsilon}{}^0_{\alpha(s+1),\:\beta(s)}

+\frac{s}{3}h^\gamma{}_\beta\stackrel{-}{\epsilon}{}^0_{\alpha(s+1),\:\beta(s-1)\:\gamma}
-(s+1)h^\gamma{}_\alpha\stackrel{-}{\epsilon}{}^0_{\gamma\:\alpha(s),\:\beta(s)}
\\
\\
\dps -
\frac{2s}{3}(s+1)h^\gamma{}_\alpha\stackrel{-}{\epsilon}{}^0_{\alpha(s)\:\beta,\:\beta(s-1)\:\gamma}
+2(s+2)h^\gamma{}_\beta\stackrel{-}{\epsilon}{}^1_{\gamma\:\alpha(s+1),\:\beta(s-1)}
\\
\\
\dps+
\frac{1}{2}(s+2)(s-1)h^\gamma{}_\beta\stackrel{-}{\epsilon}{}^1_{\alpha(s+1)\:\beta,\:\beta(s-2)\:\gamma}
-(s+2)(s+1)h^\gamma{}_\alpha\stackrel{-}{\epsilon}{}^1_{\gamma\:\alpha(s)\:\beta,\:\beta(s-1)}
\\
\\
\dps-\frac{1}{4}(s+2)(s+1)(s-1)h^\gamma{}_\alpha\stackrel{-}{\epsilon}{}^1_{\alpha(s)\:\beta(2),\:\beta(s-2)\:\gamma}\;.

\ea
\ee
According to (\ref{taus}) and (\ref{taus2}), the linearized higher
spin curvature $\stackrel{-}{r}{}^0_{\alpha(s+1),\:\beta(s)}$ has the
analogous component form.

The extra field $\omega^1$ is guaranteed to fall out from the
equations (\ref{EM}) as a consequence of the extra field
decoupling condition (\ref{exdc}). However one can check this fact
directly. To this end let us rewrite (\ref{EM}) as
$\hat{E}\bar{r}^0 =0$ (and $\hat{E}r^0 =0$ for the conjugated
field). Then, according to (\ref{taus}) the extra field enters the
equation of motion through $\hat{E}{\cal T}^-\bar{\omega}^1$ and is
required to vanish. To see this let us note that

\be
\ba{l}
\dps(\hat{E}{\cal
T}^-\stackrel{-}{\omega}{}^1)_{\alpha(s+1),\:\beta(s)}\;\delta\omega^{0\;\alpha(s+1),\:\beta(s)}\;\sim
\\
\\
\dps \stackrel{-}{\omega}{}^1_{\alpha(s+2),\:\beta(s-1)}\;({\cal
T}^+\hat{E}\delta\omega^0)^{\alpha(s+2),\:\beta(s-1)}=0\;,

\ea
\ee
The last line is easy to prove because of a simple form of the operator
${\cal T}^+$.

It should be stressed that the fermionic action (\ref{act})
generate the dynamics of a spin $s+3/2$ field which differs from
that of \cite{vf} only in the form of description of higher spin
fields. Following \cite{vf}, one can calculate the number of
physical degrees of freedom $N^s$:

\be
N^s=2(s+2)
\ee
Since Dirac spinor in three dimensions has two components, we
conclude that $N^s$ is equal to the dimensionality of the
irreducible $\gamma$-transverse rank $s+1$  tensor-spinor
representation of the little group $SO(3)$. In other words, the
action functional (\ref{act}) does describe massless fermionic
spin $s+3/2$ field in $AdS_5$. It is worth noting here that
restoring the cosmological constant $\lambda$ in the linearized
higher spin curvatures (\ref{taus}) one can easily find the flat
limit $\lambda=0$  and see that the resulting equations of motion
describe a free spin $s+3/2$ massless field in the flat space.

%----------------------------------------------------------------------------------------
\section{Conclusion}
%----------------------------------------------------------------------------------------

In this work, making use of the well-known isomorfism $o(4,2)\sim
su(2,2)$ we described fermionic higher spin fields in $AdS_5$ as
$su(2,2)$ multispinors. We demonstrated  that $\rm 5d$ higher spin
$s+3/2$ fields can be described by 1-forms
$\Omega^{\alpha(s+1)}_{\beta(s)}(x)\equiv
dx^{\underline{n}}\:\Omega_{\underline{n}\;}{}^{\alpha(s+1)}_{\beta(s)}(x)$
and their conjugates which are traceless multispinors symmetric in
the upper and lower indices.

We constructed explicitly gauge invariant free fermionic actions
which are fixed unambiguously up to an overall factor in front of
the action of a given spin. The analysis of free field equations
and the counting of physical degrees of freedom demonstrates
that the constructed fermionic action does describe a
single spin $s+3/2$ field.

The next problem to be considered is supersymmetric interactions
of $\rm 5d$ higher spin gauge fields. The results obtained in
\cite{VD5} for purely bosonic interactions along with the free
level consideration of fermions of the present work provide a good
starting point for the analysis of supersymmetric higher spin
gauge theory \cite{AV2}.

\vspace{1.5cm}

{\bf Acknowledgements}
\\
\\
I would like to express my thanks to M.A.Vasiliev for
creating favourable conditions for work. I am also grateful to him
for suggestion to consider the problem, illuminating discussions
and critical reading of the manuscript.

I benefited from useful conversations with R.R.Metsaev and I.V.Tyutin.
Valuable discussions with A.Yu.Segal on various aspects of the
higher spin problem are especially appreciated.

This work is partially supported by grant RFBR N 00-02-17956 .

\end{document}